\documentclass[aps,prd,amsfonts,nofootinbib,supercriptaddress]{revtex4}

\usepackage{amssymb,amsmath}
\usepackage{graphicx}
\usepackage{epsfig}

\newcommand{\beq}{\begin{equation}}
\newcommand{\eeq}{\end{equation}}
\newcommand{\bea}{\begin{eqnarray}}
\newcommand{\eea}{\end{eqnarray}}
\newcommand{\vc}[1]{{\textbf{#1}}}
\newcommand{\mc}[1]{\mathcal{#1}}

\begin{document}

\title{The adhesion model as a field theory for cosmological clustering}

\author{Gerasimos Rigopoulos}

\affiliation{Institut f\"ur Theoretische Physik, Philosophenweg 12,\\
Universit\"at Heidelberg, 69120 Heidelberg, Germany}

\begin{abstract}
\noindent
The adhesion model has been proposed in the past as an improvement of the Zel'dovich approximation, providing a good description of the formation of the cosmic web. We recast the model as a field theory for cosmological large scale structure, adding a stochastic force to account for power generated from very short, highly non-linear scales that is uncorrelated with the initial power spectrum. The dynamics of this Stochastic Adhesion Model (SAM) is reminiscent of the well known Kardar-Parisi-Zhang equation with the difference that the viscosity and the noise spectrum are time dependent. Choosing the viscosity proportional to the growth factor $D$ restricts the form of noise spectrum through a 1-loop renormalization argument. For this choice, the SAM field theory is renormalizable to one loop.  We comment on the suitability of this model for describing the non-linear regime of the CDM power spectrum and its utility as a relatively simple approach to cosmological clustering.

\end{abstract}

\maketitle

\section{Introduction}
The gravitational evolution of cosmological fluctuations and structure formation has been in the forefront of cosmological research for decades. Fluctuations in the early universe are small and can be understood within the framework of linear perturbation theory. A large body of work has thoroughly worked out linear perturbation theory for the CDM-Baryon-Photon mixture of early cosmic eras, driven to a large part by the promise and gradual delivery of exquisite CMB data from the WMAP and more recently the Planck satellites. This work has already understood practically all the information contained in the CMB temperature anisotropies and has pinned down the best fit cosmological model. The CMB has also shed light on primordial inflation. However, rich as it may be in information, the CMB is only a snapshot of cosmic history. The natural next step would therefore be to venture into the later stages of cosmological evolution where non-linear structures start to form.

The use of cosmological perturbation theory for tracing the evolution of CDM inhomogeneities has been explored in the past - see \cite{Bernardeau:2001qr} for a thorough review. However, the problem of large scale structure (LSS) formation is more demanding since fluctuations eventually evolve into a collection of non-linear halos with different masses and sizes, woven into the tapestry of the cosmic web visible in simulations and large scale surveys. Thus, straightforward perturbation theory is of limited use in this regime. The issue has been brought to the fore more recently given the quality and quantity of data in present and upcoming LSS surveys. These surveys will map an increasingly wide range of scales extending out to high redshifts, compelling theory to catch up on the confrontation with observation. The problem of dark energy further highlights the need for better computational technologies for cosmological clustering. Indeed, the physical properties of dark energy, whether it's static, dynamical, or even a modification of gravity, might be discernible through its effects on the formation of LSS. A major goal of many upcoming surveys, culminating in the Euclid satellite \cite{Amendola:2012ys}, is precisely the determination of these aspects of dark energy through detailed observations.

In recent years there has been a revival of interest in the further development of cosmological perturbation theory. One concrete aim has been to extend the range over which perturbative techniques can be useful, pushing into the quasi-linear and non-linear regime. After the seminal work of \cite{Crocce:2005xy, Crocce:2005xz}, which showed how conventional perturbation theory could be reorganized into something better behaved, there has been significant activity in trying to extend the validity of the theory by employing various approximation schemes in both eulerian and lagrangian space descriptions \cite{Matarrese:2007wc, Valageas:2006bi, Matsubara:2007wj, Pietroni:2008jx, Bernardeau:2011vy, Anselmi:2012cn, Crocce:2012fa, Blas:2013aba}.
All such works have assumed that the correct model of CDM on the relevant scales is that of a fully collisionless fluid. Although this is correct on scales commensurate with the mean inter-particle distance, the authors of \cite{Buchert:1997ks, Buchert:2005xj} have shown that an effective dynamical pressure arises in the coarse grained description of a collection of self-gravitating CDM particles in a cosmological setting. More recently, the authors of \cite{Baumann:2010tm} have further argued that the correct effective theory for cosmological structure formation on quasi linear scales is that of a slightly viscous fluid even though the underlying model is collisionless CDM. This picture was recently elaborated in \cite{Enqvist:2010ex, Hertzberg:2012qn, Carrasco:2012cv, Mercolli:2013bsa, Pajer:2013jj, Carroll:2013oxa, Carrasco:2013mua, Porto:2013qua, Senatore:2014via}. In this paper we utilize the idea of an effective gravitational viscosity from a different perspective to build a relatively simple stochastic field theory of structure formation, the Stochastic Adhesion Model (SAM). This is an extension of the well known Adhesion Model of \cite{adhesion1}.

In its original incarnation the adhesion model described the motion of CDM particles in the universe using Burger's equation (see \cite{adhesion1})
\beq\label{burgers}
\frac{\partial h}{\partial D} - \frac{1}{2}\left(\nabla h\right)^2=\nu\nabla^2h
\eeq
where $h$ is the velocity potential. In this model the motion of particles obeys the Zel'dovich approximation \cite{Zeldo} (see eq (\ref{Zeldovich}) below) until trajectories cross. The introduction of the visocity parameter $\nu$ ensures that shocks form at places of shell crossing which do not allow particles to overshoot, thus fixing a major deficiency of the Zel'dovich approximation. Its utility was demonstrated in \cite{adhesion2, adhesion3, adhesion4} where it was shown to produce density fields that correlate very well with the results of N-body simulations. A recent implementation of the adhesion model can be found in \cite{Hidding:2012rd}.

The original adhesion model assumed the limit $\nu\rightarrow 0_+$ leading to infinitely thin shocks. Thus the viscosity $\nu$ was seen as a mathematical device to trace out the skeleton of the cosmic web. In this paper we adopt the ideas of \cite{Baumann:2010tm} and take the view that the viscosity $\nu$ is a real attribute of a coarse grained self gravitating collection of CDM particles. It encodes short scale physics that is integrated out and its magnitude and time evolution, not calculable within SAM, can in principle be measured in N-body simulations.

Although the original adhesion model is an improvement to the Zel'dovich approximation and is able to to explain the formation of the cosmic web, the density field it produces lacks power on small scales. This is due to the absence of any modeling for processes below the scale set by the viscosity. However, as has been argued by Peebles \cite{Peebles}, small scale gravitational clustering will induce correlations on larger scales, even in the absence of initial large scale perturbations, with a characteristic $k^4$ tail as $k \rightarrow 0$. We model the effects of short scales through the addition of a stochastic noise term to (\ref{burgers}). As we will see, the requirement of producing a $k^4$ tail fixes the noise to be spatially white on large scales, i.e. with a spectrum tending to a constant at small $k$. As we further discuss, the time dependence of the noise spectrum can be fixed, given the time dependence of the viscosity, by requiring the theory to be renormalizable as the wavenumber cutoff is sent to infinity. In this work we set the viscosity proportional to the growth factor $D$, to match simulation data. The SAM is then a field theory which at tree level encapsulates a partial resummation of standard pertubation theory. It has the form of a Kardar-Parisi-Zhang equation \cite{Kardar:1986xt} with time dependent coefficients. We should add that the term stochastic adhesion model has been used in \cite{Matarrese:2001kq} which also utilized a noise term in the adhesion model in order to model the quasi-linear evolution of the baryonic fluid. Our use of the term refers to CDM alone.\footnote{The term Stochastic Adhesion Model has also been introduced in \cite{gaite}, where an argument for the necessity of a non-pertubative treatment is given. The author would like to thank J. Gaite for bringing this paper into his attention}

The outline of the paper is as follows: In section \ref{StAdMo} we obtain equations for the velocity potential $h$ and the density contrast $\delta$ by considering deviation from  the Zel'dovich approximation and discuss their physical meaning. We then formulate the equations defining the SAM. In section \ref{EFT} we cast the model in terms of an action principle and derive Feynman rules involving two types of propagators/correlation functions and a single vertex. We then discuss the time dependence of the parameters of the theory. In section \ref{renorm} we briefly discuss 1-loop corrections to the effective action and argue that the theory is renormalizable to this order. We close in section \ref{final} with some conclusions and discussion.

\section{The stochastic adhesion model}\label{StAdMo}

The fundamental equations of the CDM model specify the trajectory of each particle. For any given particle, setting as usual $\vc{x}=\vc{r}/a(t)$ where \vc{r} is the physical position, its equation of motion will be
\beq
\frac{d\vc{v}}{dt}+2H\vc{v}=-\frac{1}{a^2}\nabla_\vc{x}\phi
\eeq
where $\phi$ is the gravitational potential given by
\beq
\nabla^2_\vc{x}\phi=4 \pi G a^2\bar{\rho}_{\rm m}\,\delta
\eeq
with $\delta$ the fractional density contrast and where $\vc{v}=d\vc{x}/dt$ is the peculiar velocity. It is convenient to define new velocity and potential variables through
\beq
\vc{w}\equiv\frac{d\vc{x}}{d D}=\frac{\vc{v}}{\dot{D}}\,,\quad \psi=\frac{\phi}{4\pi G \bar{\rho}_{\rm m} a^2 D}\,,\quad
\eeq
in terms of which the equation of motion for the particle reads
\beq\label{particle eom}
\frac{d\vc{w}}{dD}+\frac{3}{2}\frac{\Omega_m}{f^2}\left(\vc{w}+\nabla_\vc{x}\psi\right)=0\,.
\eeq
The rescaled gravitational potential $\psi$ is  generated by all the particles and is given by
\beq\label{poisson}
\nabla^2_\vc{x}\psi=\frac{\delta}{D}\,,
\eeq
where $\delta$ is the density contrast given by
\beq
\delta=\frac{\rho_{\rm m}(\vc{x},D)-\bar{\rho}_{\rm m}}{\bar{\rho}_{\rm m}}
\eeq
with the density
\beq
\rho(\vc{x}, D)=\frac{m}{a^3}\sum\limits_i\delta(\vc{x}-\vc{x}_i(D))\,,
\eeq
and where the summation extends over all the particles. Here $D$ is the growing solution to
\beq
\ddot{D}+2H\dot{D}-4\pi G \bar{\rho}_{\rm m}D=0
\eeq
ie the growth factor: for linear perturbations $\delta(\vc{x},D)=D\delta (\vc{x})/D_{\rm in}$. The above equations are exact but require information about all particles in order to be solved. This is what an N-body simulation achieves.

Consider now the trajectory of one such particle coarse grained on long wavelengths. Its initial motion will be given by
\beq\label{LWVeloPot}
\vc{w}=-\nabla\psi\,,
\eeq
such that (\ref{particle eom}) gives
\beq
\frac{d\vc{w}}{dD}=\frac{\partial\vc{w}}{\partial D}+\vc{w}\cdot\nabla\vc{w}=0\,.
\eeq
The solution is of course the famous Zel'dovich approximation \cite{Zeldo}\footnote{In fact, the Zel'dovich trajectory is the long wavelength solution to General Relativity  for the trajectories of CDM particles \cite{Rampf:2012pu}.}
\beq\label{Zeldovich}
\vc{x}=\vc{q}+D\nabla\psi\,.
\eeq
Despite its simplicity, this description of the trajectory already goes some way beyond standard perturbation theory and describes the formation of the cosmic web \cite{Sahni:1995rm, Hidding:2013kka} - see also \cite{Tassev:2011ac} for another recent utilization of the Zel'dovich approximation.

As particles move closer to each other, in regions where the Zel'dovich trajectories cross, the density contrasts grow and bound structures form under the action of the local gravitational attraction between particles. This is of course not captured by the Zel'dovich approximation which itself contains no information about the local form of the gravitational potential. When this happens eq. (\ref{LWVeloPot}) breaks down. We will now explore the leading order corrections that can fix this deficiency of the Zel'dovich approximation.

For simplicity, let us ignore the generated vorticity in the velocity field and write $\vc{w}=-\nabla h$. To make analytical progress we will settle for less than the full N-body information and follow coarse grained volumes that we take to represent all the particles in the vicinity of a spacetime point - see \cite{Buchert:1997ks} for a sketch of this idea. Such volumes are taken to move with the average velocity in a given region. We can then approximate the difference between the velocity potential $h$ and the gravitational potential $\psi$ as
\beq\label{correction1}
h-\psi \simeq 0 - \nu_1 \nabla^2 \psi - \nu_2 \nabla^2 h - J_s \simeq 0- \frac{\nu_1}{D}\, \delta - \nu_2 \nabla^2 h - J_s\,,
\eeq
to leading order in a gradient expansion. As $k \rightarrow 0$, the velocity and gravitational potentials coincide and the Zel'dovich approximation is recovered. The first two correction terms on the rhs ensure that particles will stop at regions where there is a net influx and $\nabla\cdot\vc{w}=-\nabla^2h$ is negative or are overdense. The existence of such terms is also justified if the self-gravitating CDM is described on long wavelengths by an effective fluid with a small viscosity and pressure as was discussed in \cite{Baumann:2010tm}. The term $J_s$ is a stochastic component of the potential that we take to represent local, relatively fast processes related to high density small scale structures. Such processes would stochastically displace the coarse grained volume that we are following. To this approximation and transforming to Eulerian coordinates we obtain for the velocity potential $h$
\beq\label{KPZ1}
\frac{\partial h}{\partial D} - \frac{1}{2}\left(\nabla h\right)^2=\frac{\nu_1}{D}\,\delta+\nu_2\nabla^2h +J_{\rm s}
\eeq
Note that we have absorbed the (almost constant for $\Lambda$CDM) $\frac{3}{2}\frac{\Omega_m}{f^2}$ factor into $\nu$ and $J_s$. With eq. (\ref{KPZ1}) dictating the motion of particles, the equation for the density contrast $\delta$ will now obey a modified continuity equation
\beq\label{Den1}
\frac{\partial \delta}{\partial D}-\nabla h\cdot\nabla\delta-\left(1+\delta\right)\nabla^2h=\nu_3\nabla^2\delta + \tilde{J}_s
\eeq
The l.h.s. of this equation evolves the density of particles like the inverse of the infinitesimal volume element following the flow of the fluid, which is the standard physical content of the continuity equation. However, the stochastic component of the velocity equation (\ref{KPZ1}) will necessarily induce a diffusive term that we included on the r.h.s.. The physical basis for such a term is that even if there is no bulk flow, the random walk created by $J_s$ in (\ref{KPZ1}) will result in diffusion whenever there is a difference in density gradients. A stochastic force term $\tilde{J}_s$ is also expected to arise from renormalization. Thus, we have approximated the motion of particles in the complex gravitational field of small scale non-linearities with a non-linear Langevin-type equation with viscosity ie a stochastically driven hydrodynamical theory.

The system of stochastic equations (\ref{KPZ1}) and (\ref{Den1}) can be formulated in terms of an action functional and a path integral, allowing for a perturbative expansion in terms of familiar Feynman diagrams to be constructed. We will explore this route for the full system of (\ref{KPZ1}) and (\ref{Den1}) in a forthcoming publication. In the present work we will first employ a simplification that decouples the velocity from the density: we make the replacement
$\nu_1\delta/D \rightarrow \nu_1\nabla^2 h$
in (\ref{KPZ1}) to obtain
\beq\label{KPZ2}
\frac{\partial h}{\partial D}=\nu\nabla^2h + \frac{1}{2}\left(\nabla h\right)^2+ J_{\rm s}
\eeq
with $\nu=\nu_1+\nu_2$. This decouples the velocity equation from the density contrast $\delta$. To proceed, we will take the noise $J_s$ to have temporally white correlations
\beq
\langle J_{\rm s}(\vc{x}_1,D_1)J_{\rm s}(\vc{x}_2,D_2) \rangle = \Delta(\vc{x}_1-\vc{x}_2; D_1)\,\delta^D\!(D_1-D_2)\,.
\eeq
We will also add a further source term term imposing the initial conditions at the initial time $D^{\rm in}$: $J_{\rm s} \rightarrow J=J_{\rm s} + J_{\rm in}$ with
\beq
\langle{J}_{\rm in}(\vc{x}_1,D_1){J}_{\rm in}(\vc{x}_2,D_2)\rangle=\int\frac{d^3k}{(2\pi)^3}P_{\phi_{\rm in}}(k)\,e^{i\vc{k}(\vc{x}_1-\vc{x}_2)} \,\delta^D\!(D-D^{\rm in})\,\delta^D\!(D'-D^{\rm in})
\eeq
where $P_{\phi_{\rm in}}(k)$ is the power-spectrum of the initial gravitational potential. Thus our basic dynamical equation becomes
\beq\label{KPZ2}
\frac{\partial h}{\partial D}=\nu\nabla^2h + \frac{1}{2}\left(\nabla h\right)^2+ J
\eeq
with
\beq
\langle J(k,D) J(k',D')\rangle= \mc{N}(k,D)\delta(D-D')\delta(\vc{k}+\vc{k}')(2\pi)^3
\eeq
where
\beq\label{N}
\mc{N}(k,D)=P_{\phi_{\rm in}}(k)\,\delta(D-D_{\rm in})+\Delta(k,D)
\eeq
Equation (\ref{KPZ2}) is a stochastic extension of the adhesion model, first introduced in \cite{adhesion1}. A similar equation was also used in \cite{Matarrese:2001kq} to describe the dynamics of baryonic matter where the viscosity and noise represented short scale baryonic physics (see also \cite{Berera:1994sc}). Equation (\ref{KPZ2}) is a time dependent Kardar-Parisi-Zhang equation \cite{Kardar:1986xt}.

\section{Action principle and Feynman rules}\label{EFT}

We now cast the solution to equation (\ref{KPZ2}) in the form of a path integral that will allow for a diagrammatic expansion.\footnote{A similar path integral in the context of inflation was discussed in \cite{Garbrecht:2013coa}.} Both the noise term and the initial conditions encoded in $J=J_{\rm s} +J_{\rm in}$ are stochastic quantities and the expectation value of any observable $\mathcal{O}[h]$ with respect to them can be written as
\beq
\label{func:exp:stoch}
\langle \mathcal{O}[h] \rangle = \int D\!J\,{\rm e}^{-\frac{1}{2}\int \!\! J \mc{N}^{-1} J} \int D\!h\,\mathcal{O}[h]\,\delta\!\left(\partial_Dh-\nu\nabla^2h - \frac{1}{2}\left(\nabla h\right)^2- J\right)\,.
\eeq
By expressing the delta functional as a functional ``Fourier transform'' with the aid of an auxiliary field $\psi$ and performing the Gaussian $J$ integral we obtain
\beq\label{GenFunc}
\langle \mathcal{O}[h] \rangle=\int D\!hD\!\psi\, \mathcal{O}[h] \,{\rm e}^{{\rm i}\int \!\! dx \,\left[ \psi\left( \partial_Dh-\nu\nabla^2h - \frac{1}{2}\left(\nabla h\right)^2\right) + {\rm i} \psi\mc{N}\psi\right]}\,.
\eeq
It should be noted that formulations in terms of a path integral similar to (\ref{GenFunc}), but only including the initial conditions in the source term, were given in \cite{Valageas:2006bi, Matarrese:2007wc}.

\begin{figure}[t!]

\begin{center}
\parbox{4.1cm}
{
\center
\vskip2.1cm
\epsfig{file=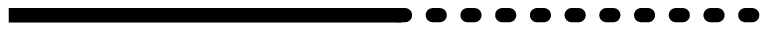,scale=0.4}

\vskip1.1cm
$G^R_k(D,D')$
}
\parbox{4.1cm}
{
\center
\vskip2.1cm
\epsfig{file=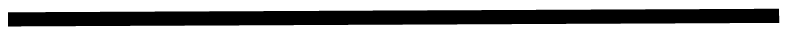,scale=0.4}

\vskip1.1cm
$C_k(D,D')$
}
\parbox{4.1cm}
{
\center
\vskip1.2cm
\epsfig{file=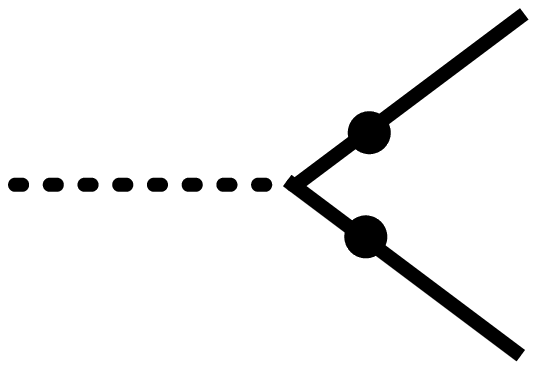,scale=0.5}
\vskip0.2cm
$\frac{i}{2}\,\left(\vc{q}_1\cdot\vc{q}_2\right)\,\delta\!\left(\vc{k}+\vc{q}_1+\vc{q}_2\right)$
}
\end{center}
\caption{The Feynman rules corresponding to the action (\ref{action1}). There are two types of ``propagator'' corresponding to the causal structure of the theory (\ref{action1}): the retarded Green function $G_k^R(D,D')$ and the correlator $C_k(D,D')$. The dots in the vertex correspond to multiplication by the wavenumber of the line. As usual, wavenumber conservation applies and internal momenta and vertex times are integrated over.}\label{Feynman}
\end{figure}

To proceed it is convenient to symmetrize the quadratic part of the action in (\ref{GenFunc}). The stochastic theory is then described by the following action
\beq\label{action1}
S=\frac{1}{2}\int dD\frac{d^3k}{(2\pi)^3}\,\left[ \left(\begin{smallmatrix}h_{\vc{k}}\,, &\psi_{\vc{k}} \end{smallmatrix}\right)\left(\begin{smallmatrix}0 & -\partial_D+\nu k^2\\ \partial_D +\nu k^2 & {\rm i}\mc{N}(k,D)\end{smallmatrix}\right)
\left(\begin{smallmatrix}h_{-\vc{k}} \\ \psi_{-\vc{k}} \end{smallmatrix}\right) + \mc{L}_{\rm int}\right]
\eeq
where
\beq\label{intraction}
\mc{L}_{\rm int}=\int \frac{d^3q_1}{(2\pi)^3}\frac{d^3q_2}{(2\pi)^3}\,\,\left(\vc{q}_1\cdot\vc{q}_2\right)\,\,\psi_{\vc{k}}h_{\vc{q}_1}h_{\vc{q}_2}\,\delta(\vc{k}+\vc{q}_1+\vc{q}_2)
\eeq
is the interaction vertex. The corresponding Feynman rules are shown in figure \ref{Feynman}.
The free correlation functions are determined as the functional and matrix inverse of the quadratic operator in (\ref{action1}) which is found to be:
\beq\label{propagators}
\left(\begin{smallmatrix}\langle h_\vc{k}(D) h^*_\vc{k}(D')\rangle & \langle h_\vc{k}(D)\psi^*_\vc{k}(D')\rangle \\ \langle\psi_\vc{k}(D)h^*_\vc{k}(D')\rangle & \langle\psi_\vc{k}(D)\psi_\vc{k}(D')\rangle\end{smallmatrix}\right)\equiv-{\rm i}\left(\begin{smallmatrix}0 & -\partial_D+\nu k^2\\ \partial_D +\nu k^2 & {\rm i}\mc{N}(k,D)\end{smallmatrix}\right)^{-1}\!\!\delta(D-D')=
\left(\begin{smallmatrix}
C_\vc{k}(D,D')&-{\rm i}G_\vc{k}^R(D,D')\\
-{\rm i}G_\vc{k}^A(D,D')&0\end{smallmatrix}\right).
\eeq
Here $G_\vc{k}^{(R,A)}(D,D')$ are the retarded and advanced Green functions for the operator $\partial_D+\nu k^2$
\beq
G^R(D,D')=G^A(D',D)=
{\rm e}^{-k^2\int_{D'}^D\!\!\nu(\eta)d\eta}\,\,\Theta(D-D')\,,
\eeq
and $C_\vc{k}(D,D')$ is the 2-point correlation function of $h$
\beq\label{corr}
\langle h_\vc{k}(D)h^*_\vc{k}(D')\rangle \equiv C_\vc{k}(D,D')= \int\limits_{0}^{+\infty} d\eta \,\,G_\vc{k}^R(D,\eta)\mc{N}(k,\eta)G_\vc{k}^A(\eta,D')\,.
\eeq

We will now write
\beq
\nu(D)=c_v^2f(D)
\eeq
where $0\leq f \lesssim \mc{O}(1)$\footnote{We have taken the growth factor $0< D \leq 1$ } is a function that encodes the time dependence and $c_v$ a constant with dimensions of length determining the scale of the effective ``viscosity'' of the system. One then obtains for the $h$ correlation function
\beq\label{h-correlation}
C_\vc{k}(D,D')=e^{-c_v^2k^2\int_{D'}^D\!\!f(\sigma) d\sigma}\left[P_{\phi_{\rm in}}(k)\,e^{-2c_v^2k^2\int_0^{D'}\!\!f(\sigma)d\sigma} + \int_0^{D'}\!\!\!\!d\eta \, \Delta(k,\eta)\,e^{-2c_v^2k^2\int_\eta^{D'}\!\!f(\sigma)d\sigma} \right]
\eeq
where we have set $D_{\rm in}\rightarrow 0$, assumed $D'<D$ and used (\ref{N}). We remind the reader that $\Delta(k,D)$ is the spectrum of the stochastic noise and $P_{\phi_{\rm in}}(k)$ is the initial power spectrum for the gravitational potential.

The form of the correlation function (\ref{h-correlation}) is suggestive. The viscosity $c_v^2$ leads to a gradual decay of the initial power, while the noise induced second term creates power even if it was absent initially. Indeed this is what happens in gravitational clustering: N-body simulations exhibit an effective dissipative behaviour which gradually decorrelates perturbations from their initial conditions, an effect evident in the decay of the Baryon Acoustic Oscillations \cite{Crocce:2007dt, Rasera:2013xfa} and which can be interpreted as loss of initial power. On the other hand, small scale gravitational clustering generates new correlations that make up for the loss of initial power and reshape the small scale spectrum. This generated component scales with a characteristic asymptotic $k^4$ dependence on large scales \cite{Peebles}. Thus, the final non-linear spectrum can be broken up into two pieces whose behaviour can be extracted from numerical N-body simulations, see eg \cite{Rasera:2013xfa}. In order to match this behaviour the viscosity term must be given a linear dependence on the growth factor
\beq
f(D)=D
\eeq
or
\beq\label{h-correlation2}
C_\vc{k}(D,D')=P_{\phi_{\rm in}}(k)e^{-c_v^2k^2\frac{D^2+D'^2}{2}}
+e^{-c_v^2k^2\frac{D^2+D'^2}{2}} \int_0^{D'} \!\!\!d\eta\,\, \Delta(k,\eta)\,e^{c_v^2k^2\eta^2}
\eeq

We must now determine the function $\Delta(k,\eta)$. In general this can be obtained from a matching calculation within a more fundamental description or measured from an N-body simulation. Here we will make a few general statements and postpone a more accurate determination for future work. As we will argue below and in the next section, a natural ansatz for the noise spectrum is
\beq
\Delta(k,D)=\Delta_1(k)D+\Delta_3(k) D^3\,.
\eeq
although more complicated functions could in principle be chosen, see section \ref{final}. The functions $\Delta_1(k)$ and $\Delta_3(k)$ should have the following properties: On large scales $\lim\limits_{k\rightarrow 0}\Delta_1(k)=\Delta_1$, $\lim\limits_{k\rightarrow 0}\Delta_3(k)=\Delta_3$, ie the noise should tend towards a scale independent amplitude at large scales. We choose $\Delta_1>\Delta_3$ i.e. the $D$ term to dominate over the $D^3$ term on large scales. On small scales we set $\Delta_1(k\gg 1/c_vD)\rightarrow 0$ sufficiently fast such that $\Delta_3(k)$ dominates in this regime. The characteristic length scale here is set by $c_v$. This distinction is of course time dependent and we will always be referring to scales for which, given a value of $D$, either $c_v k D\gtrless 1$. Our choice for the noise spectrum gives
\bea\label{h-correlation2}
C_\vc{k}(D,D')&=&P_{\phi_{\rm in}}(k)e^{-c_v^2k^2\frac{D^2+D'^2}{2}}+
\frac{\Delta_1(k)}{2c_v^2k^2}\,\left(1-e^{-c_v^2k^2D'^2}\right)e^{-c_v^2k^2(D^2-D'^2)}\nonumber\\
&&+\frac{\Delta_3(k)}{2c_v^4k^4}\,\left(c_v^2k^2D'^2+e^{-c_v^2k^2D'^2}-1\right)e^{-c_v^2k^2(D^2-D'^2)}
\eea

We are now in a position to determine the density contrast in SAM. It obeys
\beq\label{Den2}
\frac{\partial \delta}{\partial D}+\nu_3\nabla^2\delta -\nabla h\cdot\nabla\delta-\delta\nabla^2h=\nabla^2h
\eeq
which can be solved perturbatively with $\nabla^2h$ treated as an independent source with its statistics already determined. Note that there is no need for a noise term in this approximation as $\delta$ is taken to be entirely determined from $\nabla^2 h$.\footnote{This is only an approximation within SAM and not generically true, consult (\ref{KPZ1}) and (\ref{Den1})} The tree-level 2-point function of $\delta$ will then read
\beq
\langle \delta_\vc{k}(D)\delta_{\vc{k}'}(D)\rangle=k^4\,\delta(\vc{k}+\vc{k}')\,\int\limits_0^{+\infty} d\eta d\eta'  \, G(D,\eta) \, C_{\vc{k}}(\eta,\eta')\,G(D,\eta')
\eeq
where
\beq
G(\eta,\eta')=
{\rm e}^{-k^2\int_{\eta'}^\eta\!\!\nu_3(\eta)d\eta}\,\,\Theta(\eta-\eta')\,,
\eeq
is the retarded Green function for (\ref{Den1}). Choosing $\nu_3=\nu$ and using (\ref{h-correlation2}) we obtain
\bea\label{delta-correlation}
\langle \delta_\vc{k}(D)\delta^*_\vc{k}(D')\rangle =&&  k^4 P_{\phi_{\rm in}}(k)\,D'^2\,e^{-c_v^2k^2\frac{D^2+D'^2}{2}}\nonumber\\
&+&k^4\Delta_1(k)\,\frac{ D'}{2c_v^3k^3}\left( F(c_vkD')-c_vkD'e^{-c_v^2k^2D'^2}\right) e^{-c_v^2k^2\frac{D^2-D'^2}{2}}\nonumber\\
&+&k^4\Delta_3(k)\frac{D'}{8c_v^5k^5}\,\left[\left(2+4e^{-c_v^2k^2D'^2}\right)c_vkD'-6F(c_vkD')\right]e^{-c_v^2k^2\frac{D^2-D'^2}{2}}
\eea
The function $F(x)$ is the Dawson integral \cite{Dawson} with the asymptotic behavior
\beq
F(x)\simeq x\,,\quad x\ll 1; \quad F(x)\simeq\frac{1}{2x}\,,\quad x\gg 1\,.
\eeq
The noise therefore induces the following behavior for the density perturbations: At equal times $D=D'$ and for $kc_vD\ll 1$ we obtain
\beq\label{delta-correlation2}
\lim\limits_{k\rightarrow 0}\langle \delta_\vc{k}(D)\delta^*_\vc{k}(D)\rangle = k^4 P_{\phi_{\rm in}}(k)\,D^2\,e^{-c_v^2k^2D^2}
+k^4\Delta_1\,\frac{ D^4}{6}+k^4\Delta_3\,\frac{ D^6}{20}
\eeq
whereas for $c_v k D\gg 1$
\beq\label{delta-correlation3}
\langle \delta_\vc{k}(D)\delta^*_\vc{k}(D)\rangle \simeq \Delta_3(k)\frac{D^2}{4c_v^4}
\eeq

The different terms in equations (\ref{delta-correlation}) and (\ref{delta-correlation2}) have a clear physical interpretation: The term depending on the initial conditions $P_{\phi_{\rm in}}(k)$ is exponentially suppressed at small scales. This is the effect of the diffusive dynamics and indicates the gradual loss of initial power that can be interpreted as a decorrelation of the density from the initial conditions which are forgotten on length scales set by $c_v^2$. The noise terms represent power that is not directly correlated with the initial conditions but rather arises from small scale processes, modeled here as stochastic noise. The necessity of the noise term becomes evident from the fact that small scale motions of particles produce a spectrum of fluctuations on longer wavelengths even in the absence of initial power on those scales. The established spectrum is controlled by the choice of $\Delta_1(k)$. The $\Delta_1(k)D$ term produces a spectrum that is constant in time on short scales. The $\Delta_3(k)D^3$ term allows for time dependence of the noise-induced spectrum on such scales. The specific cubic time dependence is suggested by a renormalization argument that we will discuss in the next section.

The value of $c_v$ as well as the form of $\Delta_1(k)$  and $\Delta_3(k)$ can be extracted from simulations. A well known argument \cite{Peebles} shows that the second term in (\ref{delta-correlation2}) is the generic outcome of short scale gravitational interactions implying that the noise must be chosen spatially white on long wavelengths. Furthermore, the results of \cite{Rasera:2013xfa} suggest a viscosity parameter $c_v$ which takes values in the range
\beq
c_v \sim 8-10\,{\rm Mpc/h} \,.
\eeq
The viscosity parameter from \cite{Rasera:2013xfa} is scale dependent, a fact easily incorporated in the above considerations.

\section{On renormalization at one loop}\label{renorm}

In this section we briefly discuss the effect of short scale loop corrections on the field theory (\ref{action1}), postponing a more complete treatment for future work \cite{Florian}. We should stress that there is not necessarily an issue of UV divergences that needs to be addressed here; the UV behavior is determined by the exact form of $\Delta(k,D)$ and given the diffusive nature of the system it is unlikely that UV divergences will arise for reasonable forms of $\Delta(k,D)$. What we will focus on is the impact of fluctuations near a possible cutoff $L$ and the renormalization of the long wavelength sector of the theory as this cutoff is changed. We assume that for any wavenumber $k$ of interest $k \ll L$. Following a Wilsonian approach\footnote{See e.g.  \cite{Altland:2010si} for a textbook discussion of Wilsonian renormalization in a time dependent setting.} we will integrate out fluctuations with wavenumbers in a thin wavenumber slice $e^{-l}L\leqslant q\leqslant L$ with $0<l\ll 1$ and investigate the type of terms that arise in the effective action from this operation.

\begin{figure}[t!]
\begin{center}
\parbox{4.1cm}
{
\center
\vskip2.1cm
\epsfig{file=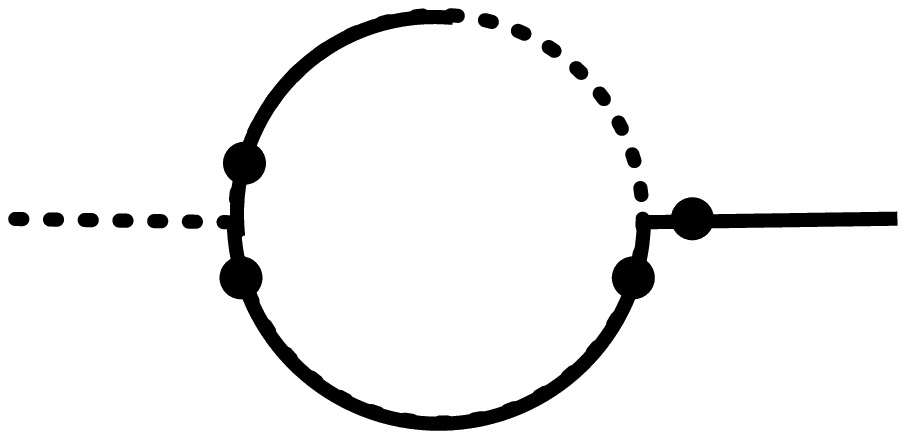,scale=0.4}
\vskip0.5cm
(A)
}
\hskip 1cm
\parbox{4.1cm}
{
\center
\vskip2.1cm
\epsfig{file=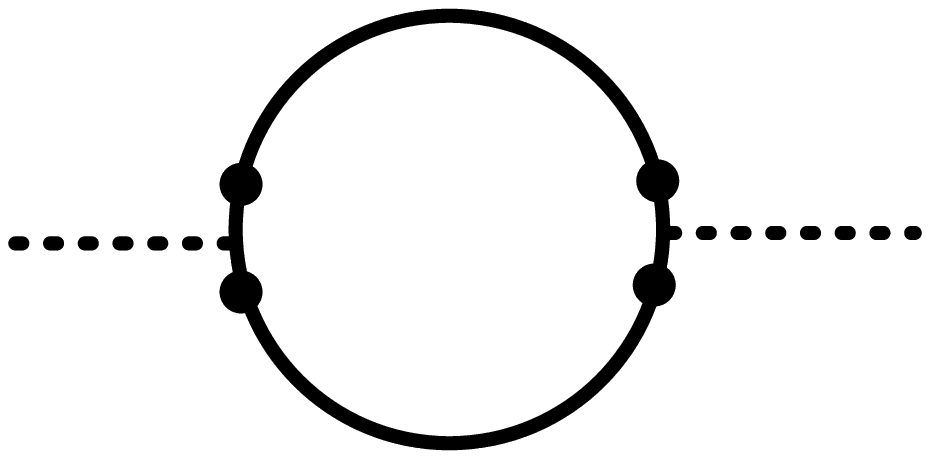,scale=0.4}
\vskip0.5cm
(B)}

\end{center}
\caption{The one loop diagrams that renormalize the action. Diagram (A) contributes to the ``$\psi-h$'' part of the action while diagram (B) adds a contribution to the ``$\psi-\psi$'' part, renormalizing the noise.}\label{loops}
\end{figure}

To begin with, the two loops that will renormalize the free part of the effective action are shown in figure \ref{loops}. Note that no propagators are attached to the external legs of the diagrams making them ``self-energy'' graphs. As we will see, in the limit of the internal momentum $q\sim L \gg k$ diagram (A) contributes a $k^2$ term and hence renormalizes the viscosity $\nu(D)$. Diagram (B) renormalizes the noise spectrum $\Delta(k,D)$. The important thing to note in this time dependent setting is that the loops will in general shift not only the overall magnitudes of $\nu$ and $\Delta(k,D)$ but will also generate new time dependent terms. However, the time dependence of $\nu$ has already been fixed: $\nu=c_v^2D$. Requiring that UV fluctuations do not spoil this linear dependence in $D$ essentially fixes the short wavelength part of $\Delta(k,D)\propto D^3$ as we will now see.

Diagram (A) of figure \ref{loops} contains the momentum integral
\beq\label{I1-integral}
I_{1\vc{k}}(D,D')=-4i\left(\frac{i}{2}\right)^2\int\frac{d^3q}{(2\pi)^3}G^R_{q_+}(D,D')C_{q_-}(D,D')
\left(\vc{k}\cdot\vc{q}_{-}\right)\left(\vc{q}_+\cdot\vc{q}_{-}\right)
\eeq
where we have kept track of the external momentum $\vc{k}$ attached to the external legs, included a combinatorial factor of 4 and defined $\vc{q}_+=\vc{q}+\frac{\vc{k}}{2}$ and $\vc{q}_-=\vc{q}-\frac{\vc{k}}{2}$. The integral $I(D,D')$ generically contributes a non-local-in-time term to the $\psi-h$ part of the effective action. As mentioned above, we will be interested in doing the integral over a thin wavenumber shell at the cutoff $L$ so we use only the $\Delta_3(q)$ part of $C_{q_-}(D,D')$, see (\ref{h-correlation2}). Since the scales of interest are far above the cutoff we can work to leading order in $k/q$, effectively taking $q\rightarrow \infty$. In this case we see that the only contribution to the integral comes from times\footnote{More precisely the only contribution comes from  $|D-D'|<\frac{1}{c_v q}\rightarrow 0$ } $D\simeq D'$, making the contribution to the effective action essentially local. We can thus integrate (\ref{I1-integral}) over $D'$ up to $D$ to obtain a local contribution to the effective action
\bea
i\Delta S_1&\sim& \frac{i}{2}\int\frac{d^3q}{(2\pi)^3}  \left(\vc{k}\cdot\vc{q}_{-}\right)\left(\vc{q}_+\cdot\vc{q}_{-}\right)\frac{\Delta_3(q_-)}{2c_v^4q^4}\left(D+2\,e^{-c_v^2q^2D^2}D-3F(c_vqD)\right)\nonumber\\
&\simeq& -Dk^2 \int\frac{d^3q}{(2\pi)^3}\frac{\Delta_3(q_-)}{4 c_v^4 q^2}
\eea
which has the correct time dependence to be absorbed into a renormalization of $c^2_v$. Note that had we included a power greater than $3$ in the noise, we would have obtained a $D^n$ viscosity term with $n>1$. Similarly, had we not suppressed $\Delta_1(k)$ in the UV we would have gotten a constant term for the viscosity which would again spoil the linear $D$ dependence.

Let us now look at the loop diagram (B). The relevant momentum integral reads
\bea
I_{2}(D,D')&=&-\frac{1}{4} \int\frac{d^3q}{(2\pi)^3} q^4 C_q(D,D')^2\nonumber\\
&\simeq& -\frac{1}{4} \int\frac{d^3q}{(2\pi)^3} q^4  \left(\frac{\Delta}{2c^4q^4}\right)^2\left(c_v^2k^2D'^2+e^{-c_v^2k^2D'^2}-1\right)^2e^{-2c_v^2k^2(D^2-D'^2)}
\eea
The arguments pertaining to diagram (A) also apply here and in the large $q$ limit we can consider this diagram to give an essentially local-in-time contribution to the $\psi - \psi$ part of the effective action, ie renormalizing the noise spectrum. After integrating over $D'$ we find
\beq
i\Delta S_2 \sim -\frac{D^3}{64c_v}\int\frac{d^3q}{(2\pi)^3}q^3\Delta_3^2(q)+\frac{3D}{256c_v^3}\int\frac{d^3q}{(2\pi)^3}q\Delta_3^2(q)\,.
\eeq
Again the time dependence of the noise spectrum is preserved and the dependence on the short scale fluctuations can be absorbed in the constants $\Delta_1$ and $\Delta_3$.

\section{Conclusions and discussion}\label{final}

In this work we have formulated a stochastic generalization of the adhesion model and cast it as a field theory with the corresponding Feynman rules. There are two free functions in SAM: the viscosity $\nu(D)$ and the noise spectrum $\Delta(k,D)$. These are functions that can in principle be fitted from N-body simulations. For example, measuring the decay of the correlation with the initial conditions will give $\nu(D)$ while the remaining power, including a fit to the non-linear short scale part of the spectrum, can be compensated through the action of the noise after an appropriate $\Delta(k,D)$ is chosen. for such an ``experimental'' determination we have used \cite{Rasera:2013xfa} and the choice $\nu(D)=c_v^2D$ reproduces the time dependent exponential decay of the memory of the initial conditions found in that reference. Furthermore, choosing $\Delta(k,D)=\Delta_1(k)D+\Delta_3(k)D^3$ with $\lim\limits_{k\rightarrow 0}\Delta_1(k)=\Delta_1$ and $\lim\limits_{k\rightarrow 0}\Delta_3(k)=\Delta_3$ makes the theory renormalizable to one loop. More specifically, the UV parts of the 1-loop diagrams renormalize the constants $\Delta_1$ and $\Delta_3$ without generating new time dependent terms (eg a $D^2$ term for $\nu$ or $\Delta$). Finally, since loop effects will generically shift $\Delta_1$ and $\Delta_3$ from zero, a $k^4$ tail is necessarily generated on large scales for the density spectrum (or, equivalently, a $k^0$ spectrum for the velocity potential). This is in accordance with a general argument about the influence of short scale gravitational interactions on large scales \cite{Peebles}.

It would be interesting to see how much of the non-linear power spectrum could be reproduced with SAM. For instance it is clear that $\Delta(k,D={\rm const})$ can be chosen such that it fits the non-linear spectrum down to arbitrarily short scales at one particular time but including the correct time evolution would be more involved. For the type of time dependence used in this paper, which is motivated by theoretical arguments, the time dependence for the short scale part of the spectrum is given in eq.~(\ref{delta-correlation3}). More complex but more accurate short scale fitting functions, such as the halo model of \cite{Smith:2002dz}, could in principle be used. Of course the question of the appropriate viscosity $\nu$ and the renormalizability of the theory would then have to be readdressed \cite{Florian}.

Before closing, let us note that the SAM (like the original adhesion model) is based on a decoupling of the evolution equation for the velocity from the density. This is not strictly necessary to make progress although it does simplify things considerably as it allows one to study $h$ separately. In fact, the approximation (\ref{correction1}) is rather general and results in two coupled equations, (\ref{KPZ1}) and (\ref{Den1}), for the velocity potential and the density contrast. All the methods that we described in this paper can be applied to this larger set of equations which, one would expect, is a more accurate description of the transition from the Zel'dovich regime to the non-linear regime. This larger set of equations is also in line with the Effective Field theory of Large Scale Structure (EFT of LSS) approach and our formulation can also be applied within that framework. The EFT of LSS approach introduces viscous terms that are allowed by symmetries and act as counterterms in the perturbative calculations of correlation functions \cite{Carrasco:2012cv}. Such terms also arise naturally due to the renormalization from the action of a noise term modeling small scale fluctuations \cite{Florian} (see also the discussion in \cite{Carroll:2013oxa}). Therefore, the reasoning presented in this paper seems to reproduce the type of equations that arise in the EFT of LSS, implying that the ``adhesive'' description of gravitational clustering \cite{Buchert:2005xj} has a natural affinity to the effective field theory view of structure formation put forward in \cite{Baumann:2010tm}.

Finally we would like to make a comment on the physical implications of such viscosity terms. In the adhesion model and for Gaussian, scale invariant initial conditions, the viscosity stabilizes structures that can be identified with the filaments and knots of the cosmic web, providing an accurate outline of the non-gaussian cosmic matter distribution on quasi-linear scales. It thus seems that the effective viscous terms that arise when CDM is coarse grained have physical implications for the kind of structures formed in non-linear gravitational clustering. This might suggest that these terms, along with a stochastic source providing power on smaller non-linear scales, could be used to describe gravitational clustering beyond two point functions and deeper into the non-linear regime than one might expect if they are treated only as counter terms in perturbative calculations. If this assertion can be further substantiated the numerical simulation of a SAM type theory could prove useful for reproducing the cosmic web and the transition to the non-linear regime with a smaller computational cost than a fully fledged, large scale N-body simulation. We leave the investigation of this research avenue for future work.

\section*{Acknowledgements}
This work was supported by the Deutsche Forschungsgemeinschaft through the TRR33 program ``The Dark Universe''. The author would like to thank Miguel Zumalacarregui and Florian F\"{u}hrer for comments on the manuscript.

\end{document}